\documentstyle[aps,preprint,prc,12pt]{revtex}

\begin{document}

\title{\bf Chaos vs. linear instability in the Vlasov equation:
a fractal analysis characterization  }

\author{A. Atalmi $^{1,3}$, M. Baldo$^{2,3}$, G.F. Burgio$^{2,3}$ and
A. Rapisarda $^{2,3}$}

\address{ \it $^1$ Centro Siciliano di Fisica Nucleare e
Struttura della Materia,
c.so Italia 57, I-95129 Catania, Italy}

\address{ \it $^2$ Dipartimento di Fisica Universit\'a
 di Catania, c.so Italia 57, I-95129 Catania, Italy}

\address{\it $^3$ I. N. F. N. Sezione di Catania,
c.so Italia 57, I-95129 Catania, Italy}

\date{8 September 1995}
\maketitle

%body of paper

\begin{abstract}
In this work we discuss the most recent results concerning
the Vlasov dynamics inside the spinodal region.
The chaotic behaviour which follows an initial regular
evolution is characterized through the calculation
of the fractal dimension of the distribution of the
final modes excited. The ambiguous role of the largest Lyapunov
exponent for unstable systems is also critically reviewed.

\end{abstract}

\newpage

%\narrowtext
%\begin{multicols}{2}
%togliere il commento per il preview dell'articolo finito

\bigskip
\bigskip
\bigskip

The character of the
mean-field dynamics inside the spinodal region has been recently deeply
investigated in order to understand the dynamical mechanisms leading
to nuclear multifragmentation \cite{hei93,bur92,bbr1,bbr2,jac}.
In this respect the realization of the occurrence of chaotic
behaviour in the Vlasov equation \cite{bbr1,bbr2}
has been suggested as a possible  explanation of the success
of statistical models \cite{gross,bondorf}
in explaining the experimental data.
In fact deterministic chaos seems the most natural dynamical mechanism
to fill the phase space, a general assumption made by
all statistical scenarios.
In this work, along the same line of refs. \cite{bbr1,bbr2} we want
to characterize in a more precise and clear way  the chaotic character of
 the mean field evolution in the spinodal region which follows an initial
linear behaviour.

A generic dynamical system, which evolves according to non-linear
equations, is non-integrable and
therefore displays chaotic motion if studied
for long enough time. The main signature of chaoticity is the extreme
sensitivity to the initial conditions, which can be revealed by
considering a set of trajectories, initially close in phase space,
and plotting the final values of a given dynamical variable vs.
the initial ones. If the dynamical system is chaotic, this plot
shows an irregular scattered behaviour. A typical example is the plot
of the deflection function for chaotic scattering \cite{ott,scatt}.
In ref.\cite{bbr2} this procedure is illustrated for two coupled harmonic
oscillators.

For the mean field dynamics one can consider a similar procedure.
A set of initially close trajectories inside the
spinodal region can be calculated and the amplitude of each mode
followed in time. If chaos occurs, for a given elapsed time, the final
amplitude, plotted against the initial one, should display strong
and irregular fluctuations. In the following, after a critical
review of the Lyapunov exponents previously calculated \cite{bbr1,bbr2},
we discuss scatter plots of the amplitude of the most excited modes during
the spinodal evolution \cite{jac}, which indicates the strength of the
coupling between the different modes (i.e. the non-linearity), as well
as the sensitivity to the initial conditions.
Moreover we quantify the dispersion by calculating the
corresponding fractal correlation dimension \cite{grass} as a function of
 time. We show that, within the values
of the parameters used in our calculation,
the spinodal evolution exhibits an irregular pattern for
the lowest modes after 40-50 fm/c, when the amplitude of the density
fluctuations is still a small fraction of the total average density.
This behaviour becomes common to all modes after 70-80 fm/c.
A full chaotic evolution is reached for all modes after 100 fm/c,
when primary "fragments" start to be apparent, thus validating
the assumptions of statistical models.

As done in refs. \cite{bbr1,bbr2} the Vlasov equation has been solved
numerically in a two-dimensional lattice using
the same code of ref.\cite{bur92}, but neglecting the collision integral.

We have studied a fermion gas situated on a large torus  with
periodic boundary conditions, and its size is kept constant during
the evolution. The torus sidelengths are equal to $L_x = 51~fm$ and
$L_y = 15~fm$. We divide the single particle phase space into several
small cells.
We employed in momentum space 51x51 small cells of
size $\Delta p_x = \Delta p_y = 40~ MeV/c$, while in coordinate
space $\Delta x = 0.3333~fm$ and $\Delta y = 15~fm$, $i.e.$ we have
only one big cell on the $y$-direction.
The initial local momentum distribution was assumed to be the one of
a  Fermi gas at a fixed temperature $T = 3~MeV$. We employ a local Skyrme
interaction which generates a mean field  $U[\rho] = t_0~(\rho /
\rho_0) + t_3~(\rho / \rho_0)^2$. The density $\rho$ is folded along
the $x$-direction with a gaussian of width $\mu = 0.61 fm$, in order
to give a finite range to the interaction. The parameters of the
force $t_0$ and $t_3$ have been chosen in order to reproduce
correctly the binding energy of nuclear matter at zero temperature,
and this gives $t_0 = ~-100.3~MeV$ and $t_3 = ~48~MeV$.  The
resulting EOS gives a saturation density in two dimensions equal to
$\rho_0 = ~0.55~fm^{-2}$ which corresponds to the usual
three-dimensional  Fermi momentum equal to $P_F =~260~MeV/c$.
Then a complete dynamical evolution is performed by subdividing the
total time in small time steps, each equal to $\Delta t = 0.5~fm/c$.

For more details concerning the mean field propagation on the
lattice, the reader is referred to ref.\cite{bur92}.

In refs. \cite{bbr1,bbr2} the density of the system was initially perturbed
mainly by means of small sinusoidal
waves with random noise in order to study
the response to minor change in the initial conditions. It was shown
that a very sensitive dependence on the initial conditions appears
as soon as the average density is inside the spinodal region.
We also found positive Lyapunov exponents, confirming that the dynamical
evolution is chaotic. In this respect an
important clarification must be done.
In fact it turns out that if the initial density is an exact harmonic
oscillation, then the evolution is regular for very long
times \cite{comment}.
This seems to be a very general statement valid also for molecular
dynamics approaches as far as only eigenmodes are initially excited
\cite{fisher}. However it should be stressed that one can initialize the
density according to several different prescriptions. In this respect the
exact eigenmode is quite a special case, whereas a perturbed sinusoidal
wave is a possible and even more realistic way of starting the system
in order to study its response.

A further clarification concerns the role of
 the Lyapunov exponents extracted
in our previous works. It
turns out that the  values obtained are very similar
to the growth factor  of the fastest mode
obtained in the analytical linear response \cite{jac,comment}.

The above considerations could be misleading and make the
reader think that the conclusions of refs.\cite{bbr1,bbr2} are erroneous.
Actually this is not the case and we want to discuss it further in the
following.

Let us consider a bunch of several randomly initialized trajectories
and the distance

\begin{equation}
        d(t)= |~~ |F_k(t)|_1 -  |F_k(t)|_2~~ | ~~,
\end{equation}
being $|F_k(t)|_1 $ and $|F_k(t)|_2 $ the modulus
of the strength of the $k$
mode excited for the trajectory $1$ and $2$.
In fig.1 we plot the evolution of $d(t)/d_0$, where $d_0 = d(t=0)$,
for 9 different trajectories.
The modes $n_k=3,4,5$ and  $n_k=12,13,14$ are considered at an initial
density $\rho/\rho_0 = 0.5$. We should remind
that the linear approximation to the Vlasov equation produces a dispersion
relation $\omega = \omega(n_k)$ for all modes below the cut-off, which
depends on the used interaction range. In our simulation the modes
$n_k=12,13,14$ are the fastest ones, whereas the modes $n_k=3,4,5$
evolve more slowly. In fig.1 it is evident a linear
and predictable growth up to a time which vary between 30-40 fm/c and
70-80 fm/c respectively for the
smallest and the largest modes excited in the simulations.
Figure 1 clearly shows that the time scale of the linear evolution
is limited to the first part. The latter depends on the modes.
This was already discussed in ref.\cite{bbr2} (see fig.7) but for one
simulation only.

It should be noted that this metric (eq.1)
takes into account the single contributions
of the $k$ modes to the complete evolution, whereas according to the
metric adopted previously in ref.\cite{bbr1,bbr2} the contribution
of all modes was taken on average.

{}From this simulation one can also extract the Lyapunov exponent.
In figure 1 it appears that, especially for the low $k$ modes, after the
linear regime, the distance between trajectories has a quite irregular
trend as a function of time. The average slope appears to be much larger
than the one of the linear regime, and in some cases it can display an
"up and down" behaviour. All that reflects the strong coupling among
all the modes. It is therefore meaningful to analyze separately
the two regimes and to extract the average slopes for each one of the two.
In the corresponding interval this gives an average value of the
Lyapunov exponent $\lambda$, being $d = d_0 e^{\lambda t}$.
The values extracted are reported in table 1, where the interval of time
in which the average has been taken is explicitely indicated.
In this case the values extracted
mode by mode are by far different from the
growth linear factor - also reported -
and only taking the average between all the modes excited one gets
a value which is similar to the linear growth of the largest exponent.

This result implies that the precise
value of the Lyapunov exponent is not
able to discriminate clearly between a chaotic and a regular behaviour
in the present case of an unstable system. However taking into account
this result
together with the broad power spectra of the excited modes,
found in ref.\cite{bbr1,bbr2}, and the sensitive dependence on the
initial condition leaves no room for a linear behaviour once the density
fluctuations reach an appreciable amplitude. It should be noticed that
chaotic behaviour is a generic statement adopted in the literature.
In general it does not mean a full randomness, which is the extreme
limit, but only an irregular motion. The Lyapunov exponent is one
way to quantify this irregularity and unpredictability.

A further way to quantify how chaotic is the system is by considering
final observable quantities as a function of initial one, as discussed
in the introduction. On the right-hand side of
figure 2 we plot in the case of 500 random-initialized simulations
and for $\rho/\rho_0 = 0.5$
the final modulus of the strength    $|F_k(t)|_f$ as a function of the
initial one $|F_k(t)|_i$.
To help understanding the evolution we plot aside the complete
numerical evolution for one event.
The change in the evolution from a linear
character to a chaotic one emerges clearly from the figure. This
variation is faster for the smaller modes
than for the biggest, as previously discussed. In a recent preprint
\cite{jac} similar scatter plots are presented. Our calculations were
performed independently of those results.

In order to quantify this irregular dynamics
we have calculated by means of
the Grassberger-Procaccia algorithm \cite{grass} the fractal correlation
dimension $D_2$ \cite{ott}.
That is we calculate the correlation integral
\begin{equation}
C(r) = {1\over M^2} \sum_{i,j}^M \Theta ( r - |{\bf z}_i - {\bf z}_j|)
\end{equation}
being $\Theta $ the Heaviside step function and ${\bf z}_i$ a vector whose
two components ($x_i, y_i$) are the initial
and final strengths of the modes $k$ plotted
in fig.2.  When r is small one gets
\begin{equation}
{}~~~~~~~~~~~~~~~~C(r) \cong r^{D_2} ~~~~~~~~~~~~~~~ a << r << L,
\end{equation}
being $a$ and $L$ the minimum and the maximum size of the set of points
considered.
Thefore by plotting the logarithms it is possible to extract $D_2$.
As an example we plot in fig.3
$log ~C(r) $ vs. $log~r$ for the mode $n_k=5$ at t=120 fm/c.
We considered as a good interval for the scaling behaviour satisfying
eq.(3) the one between $r_{min}= 7.5\cdot 10^{-5}$ and $r_{max}
= 1.5\cdot 10^{-3}$.
The dashed curve is the fit whose slope gives $D_2= 1.97 \pm 0.05 $.
The error one gets is of statistical nature and is related
to the limited  number of points considered.
In order to improve the accuracy
of the numerical algorithm the $y_i$
coordinate was normalized to the $x_i$.
This was necessary being the two scales quite different.

The general behaviour of $D_2$ as a function
of time is shown in fig.4 for
the modes  $n_k=5$ and $n_k=14$.
The dashed lines are to guide the eye. The figure shows, in line with
the previous discussion, that after an initial
linear dynamics which depends
on the mode and in any case is not longer than 70-80 fm/c the evolution
becomes more and more chaotic till a complete randomness. In fact $D_2=2$
is what one would expect and actually gets - we have tested our algorithm
in this sense -  when distributing random numbers on a plane.

In conclusion,  the present investigation confirms without any doubt that
after some short time which of
course depends on the parameters used - for example the range of the force
adopted - a fully chaotic behaviour occurs in the Vlasov dynamics
inside the spinodal region. This result confirms the crucial role played
by deterministic chaos in filling the phase space, a fundamental assumption
for justifying the  validity of statistical approaches. On
the other hand the
success of the latter in explaining most of the experimental data is a
confirmation of this scenario. However when the fragmentation regime is
very fast and chaos is not fully developed, some memory of the modes
initially excited can probably remain \cite{jac}.
We note in passing that a
chaotic behaviour with respect to multifragmentation
has been recently found also in other models \cite{feldmeier,aldo}.

We would like to thank Ph. Chomaz and M. Colonna for many stimulating
discussions.

\noindent

%\newpage

%\end{multicols}
%togliere il commento per il preview dell'articolo finito

%%%%%%%%%%%%%%%%%%%%%%%%%%%%%%%%%%%%%%%%%%%%%%%%%%%%%%%%%%%%%
% table
\newpage

\begin {table}

\begin{center}
\begin{tabular}{|c|c|c|c|c|}
\hline\hline
$ n_{k} $&$t_{l}~ (fm/c)$ & $t_{r}~ (fm/c)$ & $\lambda * 10^{2}~
(c/fm)$ & $\omega_{k} * 10^{2}~ (c/fm) $ \\
\hline
      3 &         25.     &    50.
& 2.3                        & 2.3  \\
      3 &         50.     & 105.
& 7.0                       &       \\
\hline
      4 & 25.             & 50. & 2.6  & 3.0  \\
      4 & 50.& 105.& 9.3 &       \\
\hline
      5 & 25.& 50. & 3.4  & 3.7  \\
      5 & 50.& 105.& 8.4 &       \\
\hline
     12 & 50.& 90. & 6.0  & 6.9   \\
     12 & 90.& 105.& 4.7  &       \\
\hline
     13 & 50.& 90. & 6.1  & 7.1  \\
     13 & 90.& 105.& 3.9  &      \\
\hline
     14 & 50.& 90. & 6.0  & 7.0  \\
     14 & 90.& 105.& 4.7  &      \\
\hline
$\leq$  24 & 50.& 105.& 5.6  & 6.3 \\
\hline \hline
\end{tabular}
\caption{For several $n_k$ modes, we
show a comparison between the Lyapunov
exponent $\lambda$ and the frequency $\omega$ predicted by the linear
response approximation. As
indicated, $\lambda$ depends on the time interval
$\Delta t = t_r - t_l$ over which the average slope is estimated.
$t_l$ and $t_r$
are reported in the second and third column. Please note the last row,
where $\lambda$ and $\omega$ are compared in the case the metric
contains the contributions of all modes summed up to $n_k = ~24$. }
\label {tab1}
\end{center}
\end{table}
%%%%%%%%%%%%%%%%%%%%%%%%%%%%%%%%%%%%%%%%%%%%%%%%%%%%%%%%%%%%%
\newpage
%%%%%%%%%%%%%%%%%%%%%%%%%%%%%%%%%%%%%%%%%%%%%%%%%%%%%%%%%%%%%
% figura 1
\begin{figure}
%\psdraft
% \begin{center}
%  \begin{turn}{270}
%\mbox{{\epsfxsize=8truecm \epsfysize=8truecm \epsfbox{dtdo3-14.ps}}}
%  \end{turn}
% \end{center}
\caption{ The time evolution of $d(t)/d_0$ is shown for different modes
and several trajectories at an initial density $\rho/\rho_0 = 0.5$.
On the left-hand side we display the behaviour
of the slowest $n_k$'s, whereas on the right-hand side the behaviour
of the fastest modes is plotted.}
\end{figure}
%%%%%%%%%%%%%%%%%%%%%%%%%%%%%%%%%%%%%%%%%%%%%%%%%%%%%%%%%%%%%
%%%%%%%%%%%%%%%%%%%%%%%%%%%%%%%%%%%%%%%%%%%%%%%%%%%%%%%%%%%%%
%figura 2
\begin{figure}
%\psdraft
% \begin{center}
%  \begin{turn}{270}
%\mbox{{\epsfxsize=8truecm \epsfysize=8truecm \epsfbox{zz_rho_5_14.ps}}}
%  \end{turn}
% \end{center}
\caption{On the left-hand side we show the time evolution of a density
profile vs. the position. The initial average density
is half normal nuclear
matter density. On the right-hand side we display scatter plots of the
Fourier spectra of a bunch of 500 trajectories. Only two modes
$n_k=5$ and $n_k=14$ are shown. See text for more details.}
\end{figure}

%%%%%%%%%%%%%%%%%%%%%%%%%%%%%%%%%%%%%%%%%%%%%%%%%%%%%%%%%%%%%
%%%%%%%%%%%%%%%%%%%%%%%%%%%%%%%%%%%%%%%%%%%%%%%%%%%%%%%%%%%%%
% figura 3
\begin{figure}
%\psdraft
% \begin{center}
%  \begin{turn}{270}
%\mbox{{\epsfxsize=8truecm \epsfysize=8truecm \epsfbox{corr_k_t.ps}}}
%  \end{turn}
% \end{center}
\caption{The log of the correlation integral C(r) is shown for $n_k=5$
at time $t= 120~fm/c$ as a function of the log of the distance parameter r
according to the definition of eq.(2).
The circles are the results of the numerical
simulations and the dashed line is a linear fit.}

\end{figure}
%%%%%%%%%%%%%%%%%%%%%%%%%%%%%%%%%%%%%%%%%%%%%%%%%%%%%%%%%%%%%
%%%%%%%%%%%%%%%%%%%%%%%%%%%%%%%%%%%%%%%%%%%%%%%%%%%%%%%%%%%%%
% figura 4
\begin{figure}
%\psdraft
% \begin{center}
%  \begin{turn}{270}
%\mbox{{\epsfxsize=8truecm \epsfysize=8truecm \epsfbox{d2.ps}}}
%  \end{turn}
% \end{center}
\caption{The time evolution of the correlation dimension $D_2$ is plotted
for the modes $n_k=5$ and $n_k=14$.}

\end{figure}
%%%%%%%%%%%%%%%%%%%%%%%%%%%%%%%%%%%%%%%%%%%%%%%%%%%%%%%%%%%%%
\end{document}